\documentclass[aps,pre,twocolumn,amsmath,amssymb,nofootinbib,floatfix,superscriptaddress,10pt]{revtex4-1}

\usepackage{amsmath,braket}
\usepackage{amssymb}
\usepackage{amsthm}
\usepackage[dvipdf]{color}
\usepackage{graphicx}
\usepackage{dcolumn}
\usepackage{bm,subfigure}
\usepackage{xcolor,colortbl}
\usepackage[normalem]{ulem}

\usepackage{amsmath,amssymb}

\DeclareMathOperator{\EX}{\mathbb{E}}

\usepackage{amsmath,amsfonts,amssymb,amsthm, bm}

\begin{document}

\title{
Lyapunov exponents explain disorder-induced polarization and soliton teleportation in a mechanical Markov system
}

\author{Will Stephenson}
\affiliation{
 Department of Physics,
  University of Michigan, Ann Arbor, 
 MI 48109-1040, USA
 }
\author{Nan Cheng}
\affiliation{
 Department of Physics,
  University of Michigan, Ann Arbor, 
 MI 48109-1040, USA
 }
 \affiliation{
 School of Physics,
  Georgia Institute of Technology, Atlanta, 
 GA 30332, USA
 }
\author{Kai Sun}
\affiliation{
 Department of Physics,
  University of Michigan, Ann Arbor, 
 MI 48109-1040, USA
 }
\author{Xiaoming Mao}
\affiliation{
 Department of Physics,
  University of Michigan, Ann Arbor, 
 MI 48109-1040, USA
 }
 \affiliation{Center for Complex Particle Systems (COMPASS), University of Michigan, Ann Arbor, USA}

\begin{abstract}
Using a mapping between spatial disorder and temporal stochasticity, 
we develop a new framework using Lyapunov exponents to explain exotic wave localization and mobility phenomena in disordered one-dimensional (1D) mechanical systems that can be constructed via a spatial analog of a Markov process, which we call ``mechanical Markov systems.''
We show that disorder induces robust polarization of zero modes (ZMs) in these mechanical Markov systems, and this phenomenon is explained using Lyapunov exponents. Remarkably, these ZMs become mobile solitons in the nonlinear regime despite the disorder-controlled localization of all other modes, and display a set of new nonlinear dynamics features including reflectionless chirality-dependent teleportation, which can also be explained using  Lyapunov exponents. 
Our results establish the Markov formalism as a powerful tool to explain and design localization and dynamics in disordered mechanical systems, opening opportunities for programmable metamaterials with novel linear and nonlinear responses.
\end{abstract}

\maketitle

\section*{Introduction}
Mapping between space and time has been central to many fronts of physics, from Thouless pumping~\cite{thouless1983quantization}, Floquet topological states~\cite{rechtsman2013photonic,li2024topological}, time crystals~\cite{wilczek2012quantum}, quantum criticality~\cite{Sondhi1997,sachdev1999quantum}, polymer---quantum mechanics mapping~\cite{gennes1968soluble}, temporal imaging~\cite{kolner1994space}, synthetic non-hermicity~\cite{xiu2023synthetically}, to AdS/CFT correspondence~\cite{maldacena1999large}. The identification of \emph{spatial order} with \emph{temporal coherence} reveals deep mathematical structures behind complex phenomena.  In light of these classical theories, it is natural to wonder: can we extend this mapping to disordered systems? 

Disorder is ubiquitous in nature, and has been increasingly utilized in engineered materials  for their robustness against structural defects and fracture \cite{curtin1997toughening,manzato2012fracture,zhang2017fiber,ritchie2021toughening}, unusual mechanical responses \cite{reid2018auxetic, hanifpour2018mechanics, rayneau2019density}, control \cite{liu2025disordered, zaiser2023disordered}, and ease of scalable manufacturing \cite{REIS201525, whitesides2002self}.
However, properties, such as wave propagation, are difficult to predict in disordered materials ~\cite{anderson1958absence,torquato2002random,Sheng26042007}.
In one dimension (1D), even infinitesimal disorder localizes all eigenstates due to Anderson localization, suppressing long-distance transport~\cite{abrahams1979scaling}, and where a given mode localizes depends sensitively on the disorder realization and is generally difficult to predict.

In this paper, utilizing a mapping between \emph{spatial disorder} and \emph{temporal stochasticity}, we discover a class of exotic wave phenomena originating from disorder.  
We report a one-dimensional (1D) Maxwell mechanical system with strong disorder where a polarization, analogous to the Kane-Lubensky topological polarization~\cite{Kane2013,Chen2014}, emerges from disorder. 
This polarization localizes a zero-frequency mode (ZM) to one end of the chain depending on the construction.  
In the nonlinear regime, this ZM is mobile throughout the system, despite the localization of all other waves under disorder.  Furthermore, this system demonstrates a set of striking effects, ranging from long-distance soliton transport that is robust against disorder to reflectionless chirality-dependent teleportation, in which the soliton either skips across a finite segment of the chain and re-emerges from a distance, or nucleates a distant, counter-propagating partner, depending on the signal’s chirality. 

Crucially, we explain these phenomena via a space-to-time mapping that identifies the 1D Maxwell mechanical system with a Markov process \cite{strogatz2000, abarbanel1993local, lyapunov1992general}.
Disorder induces a pseudo-time steady state with a well-defined Lyapunov exponent that sets this disorder-induced polarization.
This contrasts with the topological polarization extensively studied in ordered lattices where the bulk bands determine the topological winding number and thus the polarization~\cite{Kane2013}.  Although it has recently been shown that topological mechanical winding numbers can be extended to systems with disorder and aperiodicity~\cite{zhou2018topological,zhou2019topological,Lo2021}, the phenomena reported here is fundamentally different, where the polarization comes from disorder itself, without a topological winding number. 
In this case, all nonzero frequency waves are localized by disorder and do not support any winding number.  The polarization is instead induced by disorder and protected by stochasticity.  
Interestingly, disorder also causes fluctuations from this steady state in the form of “defects”, analogous to kink-antikink pairs~\cite{manton2004topological,zhou2017kink}.  
Within these defects, the Lyapunov exponent becomes positive, indicating pseudo-time chaos.
When real-time dynamics is included, the ZM becomes mobile and can traverse the system—despite the localization of all linear modes—and the defects underlie soliton teleportation.

We show that the extension of space-time mapping to disordered systems offer a realm of new correspondences between temporal stochasticity and spatial disorder. As we discuss in this paper, such correspondences demonstrate that disorder can be utilized as a design rule to program novel linear and nonlinear responses that are absent in ordered systems.
Recent advances in metamaterials revealed the rich space of properties, such as Poisson’s ratio, bandgap, and dissipation, that can be computationally engineered through disorder~\cite{yu2021engineered}.
Our work provides a distinct example in which disorder produces behaviors beyond those of ordered systems, while remaining amenable to analytic treatment. 

\section*{Model and Phenomena}\label{model}

\begin{figure*}[t!]
    \centering
    \includegraphics[width=\linewidth]{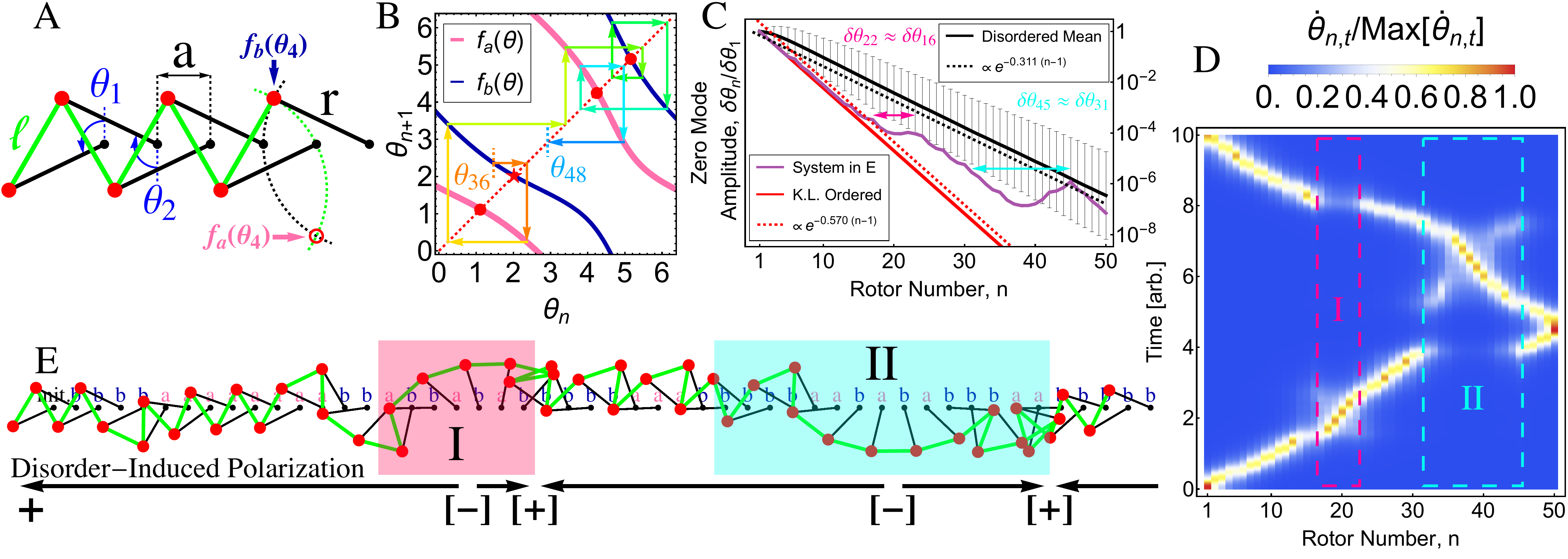}
    \caption{
    Mechanical Markov rotor chain construction and observations of ZM polarization and soliton dynamics. 
    (A) A rotor chain system with geometric parameters ($l=1, r=1, a=0.5$) such that for any possible $\theta_n$, there will always be two possible solutions for $\theta_{n+1}$ (either $f_a, f_b$,condition: $a<l<r+a/2$).
    Coordinate labeling of this system is two-periodic.
    (B) The two possible solutions for $\theta_{n+1}$ as a function of $\theta_n$ (pink/thick, purple/thin) have four fixed points (red dots, star), at which ordered systems can be constructed (panel A constructed at the star, at which $\lambda(\star) = -0.570$).
    Disordered chains are constructed by choosing between the allowed solutions at random.
    Here, this is represented via a cobweb plot for a section (rotors 36 through 48) of the system in panel E.
    (C) The geometric mean of zero mode (ZM) amplitude ($\delta\theta_n$) in $10^3$ disordered ($p_a=0.5$) chains (black, solid, error bars $= 1\sigma$) has steady-state exponential decay that is in good agreement with theory decay rate (black, dashed). 
    The left-localized Kane-Lubensky (KL) chain (red line), necessarily constructed at the one of the fixed points in (B), has a much faster decay rate (red, dashed) equal to the local Lyapunov exponent at those points (which are equal by symmetry, see SI Sec.~\ref{sym}).
    (purple) The floppy mode amplitude of a particular rotor chain, shown in panel E.
    (D) The spatiotemporal map of soliton transport through this system shows defect regions of modified soliton transport, where an outgoing (incoming) soliton travels slowly (instantly) [I] or instantly (slowly) [II]. Colors show local rotor angular velocity $\dot{\theta}_{n}$ at time $t$. 
    (E) The fifty rotor chain used in panels B-D, with solution choices ($a,b$) shown, and defect regions highlighted.
    The disorder-induced polarization is shown as black arrows, which points in the direction of increasing ZM amplitude, leading to quasi topological charges $[+]([-])$ where polarization converges (diverges). 
    }
    \label{fig:fig1}
\end{figure*}

We start from a 1D chain of rotors which we will demonstrate, is a simple mechanical Markov system (Fig.~\ref{fig:fig1}A).
The rotor chain  consists of equally spaced pivot points (separation distance $a$), around which perfectly stiff rotors of length $r$ may freely rotate \cite{Kane2013, Chen2014}.
The free ends of each rotor arm are connected to the free ends of its two adjacent rotor arms by a spring of length $l$.
The angle of each rotor arm ($\theta_n$) is measured counter-clockwise (clockwise) for every odd (even) rotor, per convention \cite{Chen2014}.

From an initial rotor with given angle $\theta_1$, we can add additional rotors so long the free ends of these two rotors are separated by a distance $l$.
Appropriate geometric parameters ($a<l<r+a/2$) ensure that there will always be two possible values of $\theta_{n+1}$ that meet the spring length constraint for any value $\theta_{n}$, allowing indefinite addition of rotors to a chain (``spinner phase'' from \cite{Chen2014}).
The possible angles of additional rotors can be found by solving the following geometric equation for $\theta_{n+1}$ as a function of $\theta_n$ (geometry shown by the dashed lines in Fig.~\ref{fig:fig1}A),

\noindent
\begin{align}
    a^2 + 2r(r+r\cos{(\theta_{n}+\theta_{n+1})} + \\ \nonumber a\sin{(\theta_n)} - a\sin{(\theta_{n+1})} ) = l^2.
\end{align}

\noindent
We call these two solutions $f_a(\theta_n)$ and $f_b(\theta_n)$, where the former is distinguished from the latter by having the lesser value mod $2\pi$ at $\theta_n=0$ (as in Fig.~\ref{fig:fig1}B).
The analytic forms of $f_a$ and $f_b$ are shown in the Supplementary Information Sec.~\ref{analytic}.

We construct disordered rotor chains by selecting an initial angle $\theta_1$ uniformly from $[0,2\pi]$, and then choosing from the two available angle solutions ($f_a, f_b$), with probabilities $p_a+p_b=1$ for each subsequent rotor, as demonstrated in a cobweb plot shown in Fig.~\ref{fig:fig1}B.
Different levels of order and disorder can be achieved by choosing between these solutions with different probabilities.
For example, the perfectly-ordered Kane-Lubensky chain \cite{Kane2013} requires $p_a=1$ or $0$.
Since this construction determines the value of a new degree of freedom ($\theta_{n+1}$) as a function of only its previous degree of freedom ($\theta_{n}$), it is exactly analogous to a Markov process, as we show later.
For this reason, we call this class of systems ``mechanical Markov systems.''

Having one more degree of freedom than constraint ($N$ rotors, $N-1$ springs), any finite-length rotor chain with free boundary conditions, like other 1D Maxwell lattices, has exactly one ZM for the entire lattice \cite{Maxwell1864}.
When a rotor chain system is spatially ordered, such that all of its angles are equal (as in Fig.~\ref{fig:fig1}A), it is well understood that this ZM must exponentially localize to one of the ends (except for the special case when geometric parameters permit a configuration where all rotors are vertical or horizontal), forming a Kane-Lubensky rotor chain with a topological edge mode \cite{Kane2013}.

Remarkably, these disordered rotor chains, where the angles are chosen stochastically between two branches, also have ZMs that  exponentially localize to one end consistently (Fig.~\ref{fig:fig1}C). 
The topological theory, which depend on long range order and momentum space, does not apply to this disordered chain.  The ZM polarization, as we discuss later, is a new emergent phenomena from disorder, which we call ``disorder-induced polarization''.  
Furthermore, one novel feature that arises due to disorder are ``kink-antikink pairs'' (Fig.~\ref{fig:fig1}E where small sections of the chain exhibit the opposite  polarization, leaving $[+]$ and $[-]$  ``quasi topological charges'' at the interfaces between sections, corresponding to localized quasi ZMs and SSSs, as we discuss in detail later.  
Unlike in ordered KL chains, where kink-antikink pairs require special construction involving defining different rest lengths~\cite{zhou2017kink}, the disordered chain here is constructed following the same rule from left to right, and these kink-antikink pairs spontaneously arise. 

It is known that in ordered rotor chains, in addition to well-understood, topologically protected  edge ZMs, there are also topologically protected zero-energy solitons ~\cite{Chen2014,zhou2017kink,Lo2021,qian2025observation, lo2021topology}. 
That is, when an rotor in an ordered chain is given some infinitesimal angular velocity, a soliton will uniformly travel through the chain, appearing as a domain wall between two oppositely polarized sub-chains \cite{Chen2014}.

Interestingly, we find that disordered rotor chains exhibit  similar soliton transport, where zero energy solitons remain mobile despite the strong disorder, despite lacking topologically protected polarization.
Moreover, the aforementioned kink-antikink pairs give rise to unique ``defect'' regions in the soliton dynamics of these disordered chains.
These regions either result in solitons traveling slowly with every individual rotor completing a nearly a full rotation (such as in Fig.~\ref{fig:fig1}D, defect I, $t\approx2$), or solitons traveling nearly instantly with every individual rotor hardly rotating (such as in Fig.~\ref{fig:fig1}D, defect II, $t\approx4$).
Note that defect II presented in Fig.~\ref{fig:fig1} is chosen to be unusually large for the sake of visual clarity.

\section*{Mapping to a Markov Process and Disorder-Induced Polarization}

\begin{figure*}[t!]
    \centering
    \includegraphics[width=\linewidth]{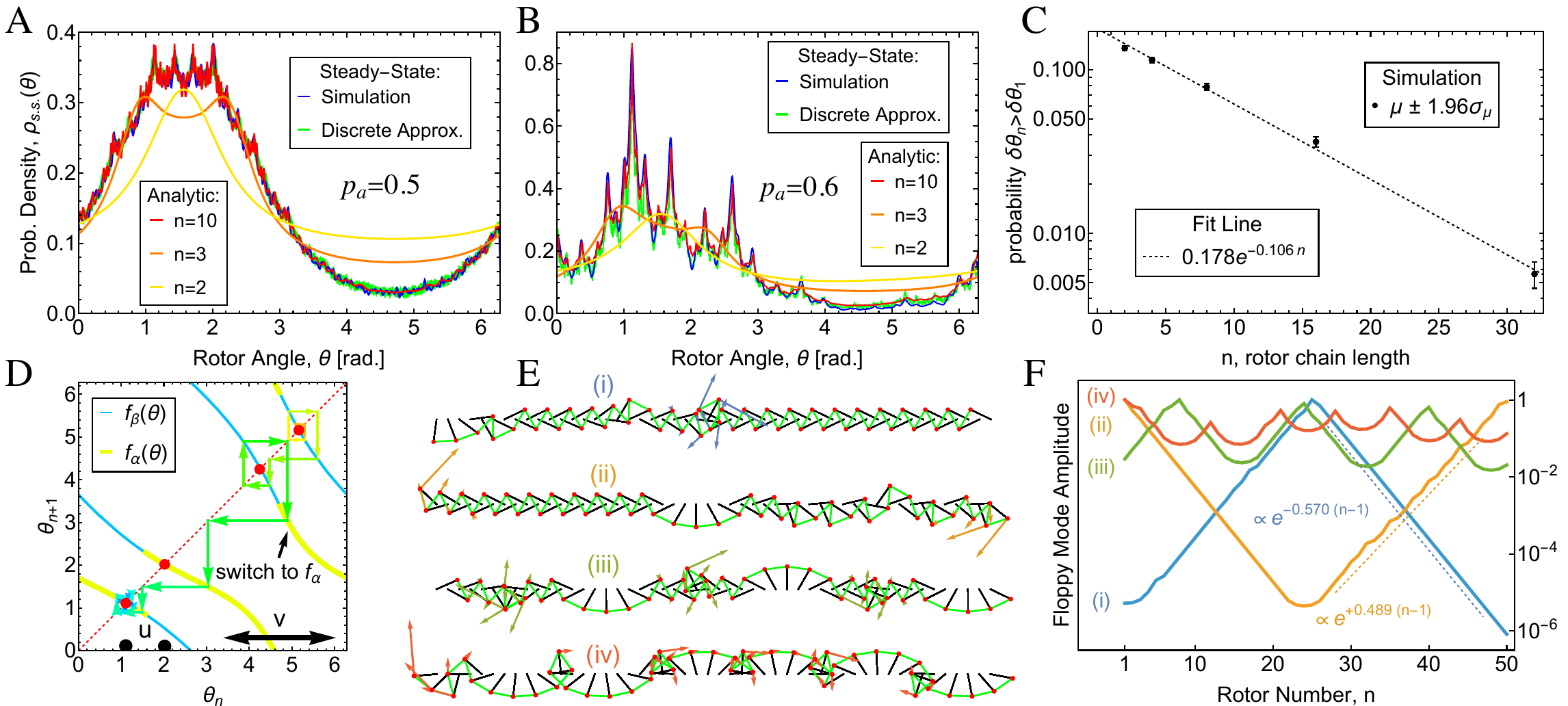}
    \caption{
    Stochastic protection of disorder-induced polarization in the 1D mechanical Markov rotor chain, and related design principles.  
    (A) Disordered rotor chains ($p_a = p_b = 0.5$), in the large system limit, reach a steady-state rotor angle probability density, in which rotor angles where the solution functions have smaller slopes ($\sim$1-2 rads.) are significantly more probable than rotor angles with large slopes ($\sim$4-5 rads.).
    Analytic results for $\rho_{10}(\theta)$ (blue, from Eq.~(\ref{eq:theoryPDF}) are in good agreement with simulation frequency (blue, mean of $10^5$ rotors) and the steady-state eigenvector a discrete Markov transition matrix of $10^4$ bins (green, from SI Sec.~\ref{disMarkov}).
    Analytic results for $\rho_{2}(\theta),\rho_{3}(\theta)$ show rapid, convergence to the steady-state limit.
    Note that analytic results shown here are actually many discrete points for computational simplicity, but could, in principle, be calculated continuously.
    (B) Choosing with other branch probabilities (such as $p_a =0.6, p_b =0.4$) results in a different $\rho_{\text{s.s}}$.
    (C) Due to the skewed probability density (induced by $p_a=0.5$), rotor chains of length $n$ ($\theta_1$ is chosen from $\rho_{\text{s.s}}$ to avoid convergence effects) are exponentially unlikely to be more floppy at position $n$ than position $1$, as shown by mean probability ($\mu$) and standard error ($\sigma_\mu$) of $2*10^4$ simulations each of rotor chains of lengths $2,4,8,16,32$.
    (D) From the two possible solutions for $\theta_{n+1}$ ($f_a, f_b$), one can create two piecewise functions of $\theta_n$ (yellow, thick: $f_\alpha$, light blue: $f_\beta$), which result in deterministic successive solutions with positive and negative Lyapunov exponents respectively (within absorbing subsets $u, v$).
    By choosing repeated solutions from $f_{\alpha (\beta)}$, ZM amplitude is guaranteed to decrease (increase).
    By choosing subsequent angle solutions (colored arrows) initially from $f_\beta$, then from $f_\alpha$ (transition marked by black arrow), one can construct a chain whose ZM initial grows and then decays.
    (E, F) By switching back and forth between these solutions branches, one can create ZMs that are localized to the center (i) or both ends (ii) of a chain, or more generally, any pattern that is a combination of exponential growth/decay (iii, iv).
    }
    \label{fig:fig2}
\end{figure*}

Consider infinitesimally activating the single  ZM of any rotor chain by changing $\theta_1$ by an amount $\delta\theta_1$ such that the new angle is $\theta_1' = \theta_1 + \delta\theta_1$. 
This process imparts no energy upon the system, requiring that all subsequent angles remain solutions of $f_a(\theta_n)$ or $f_b(\theta_n)$. 
We can then find a recursive relation for ZM amplitude ($|\delta\theta_n|$),

\noindent
\begin{align}
    \frac{\delta\theta_{n+1}}{\delta\theta_{n}}&= f_i'(\theta_n), i=a\; \text{or}\; b .
\end{align}

\noindent This relationship is exactly analogous to local Lyapunov exponents of ``discrete time'' dynamical systems \cite{strogatz2000}, where here, the local Lyapunov exponent of a given branch ($\lambda_a(\theta_n), \lambda_b(\theta_n)$) is

\noindent
\begin{align}
    \lambda_{a (b)}(\theta_n) \equiv \log{\left(\Bigg|\frac{\delta \theta_{n+1, a(b)}}{\delta \theta_n}\Bigg|\right)} = \log \left(\big\lvert f_{a(b)}'(\theta_n)\big\rvert\right).
    \label{eq:localLyap}
\end{align}

\noindent Giving the following relationship for ZM amplitude at a distant site $n$,
\begin{align}
    |\delta\theta_n| = |\delta\theta_1| \prod_{i=1}^{n-1} e^{\lambda_{a(b)}(\theta_i)}, \label{eq:relfm}
\end{align}

\noindent where the application of $\lambda_a$ or $\lambda_b$ for a given angle $\theta_i$ must match exactly with the choice of $f_a$ or $f_b$ from which $\theta_{i+1}$ was chosen from. 

To find the expected steady-state growth/decay of ZM amplitude through a rotor chain, we would need to find the expected steady-state distribution of local Lyapunov exponents, which depends on the expected steady-state distribution of rotor angles.
Given an initial uniform probability density ($\rho_1(\theta)$), the steady-state solution to the following recursion relation (averaged over $f_a, f_b$ with $p_a=p_b=0.5$) derived through probability change-of-variables (SI Sec.~\ref{rhos}) will give us $\rho_{\text{s.s.}}(\theta)$,

\noindent
\begin{align}
    \rho_{n+1}(\theta_{n+1}) &= \frac{1}{2} \left(\frac{\rho_{n}(\theta_n^a)}{\lvert f_a'(\theta_n^a) \rvert}+\frac{\rho_{n}(\theta_n^b)}{\lvert f_b'(\theta_n^b) \rvert}\right),
    \label{eq:theoryPDF}
\end{align}

\noindent where $\theta_n^{a(b)} = f_{a(b)}^{-1}(\theta_{n+1})$ for brevity.

We can then find an expected steady-state local Lyapunov exponent ($\bar{\lambda}$) by integrating over the steady-state Markov probability density (again, averaged over $f_a, f_b$ with $p_a=p_b=0.5$),

\noindent
\begin{align}
    \bar{\lambda} = \int_0^{2\pi} \frac{1}{2}\left(\lambda_a(\theta)+\lambda_b(\theta)\right)\rho_{\text{s.s.}}(\theta) d\theta.
    \label{eq:meanLyap}
\end{align}

\noindent 
The ZM will then have a steady-state limit expectation value of

\noindent
\begin{align}
    \EX(|\delta\theta_n|) = |\delta\theta_1| e^{\bar{\lambda} (n-1)}.
\end{align}

\noindent
As an alternative to using Eq.~(\ref{eq:theoryPDF}) to find $\rho_{\text{s.s.}}(\theta)$, one could approximate the steady-state probability density by finding the eigenvectors of a  discrete Markov transition matrix $\mathcal{M}$, described in SI Sec.~\ref{disMarkov}.
For the given geometry ($l=1,r=1,a=0.5)$, both the analytic steady-state probability density ($\approx\rho_{10}(\theta)$) and this discrete approximation eigenvector are in good agreement with simulation results that angles of smaller slope (which cause ZM decay) are significantly more probable (see Fig.~\ref{fig:fig2}A/B). 

Using this probability density, and Eq.~(\ref{eq:meanLyap}), we calculate a Lyapunov exponent of $\bar{\lambda}_{\text{disorder}} \simeq -0.311$, giving a mean ZM localization of $\delta\theta_n/\delta\theta_1 \simeq e^{-0.311 (n-1)}$, which is in excellent agreement with steady-state simulation results (see Fig.~\ref{fig:fig1}C).  
This exponential localization of the ZM at the initial end of the chain, due to a steady state probability distribution of configurations, defines the aforementioned ``disorder-induced polarization.''

For an ordered chain constructed exactly at a fixed point (all angles equal, $p_a = 0$ or $1$), $\bar{\lambda}$ is simply equal to the local Lyapunov exponent at that fixed point.
Here, the two stable fixed points have $\lambda_{a,b}(\theta_{\text{f.p.}}) \simeq -0.570$, giving the ordered decay rate of $\delta\theta_n/\delta\theta_1 \simeq e^{-0.570 (n-1)}$ (slopes equal because $\theta_n \rightarrow \pi-\theta_n$ and $\theta_n \rightarrow 2\pi-\theta_n$ do not physically change system, see SI Sec.~\ref{sym} for further details).
This method of linearization around only the fixed points of the transfer functions as a means of explaining the ZMs of ordered rotor chains has been shown before by Zhou et al. \cite{zhou2017kink}.

While the analysis so far has focused on the rotor chain for its familiarity, this process is in no way limited to only rotor chains system, and can be applied to any 1D mechanical Markov system.
For a certain configuration of a 1D Markov system (with $m$ different solution branches and solution choices $c_j\in[m]$), one can calculate its ZM localization by finding the local Lyapunov exponent at each degree of freedom,

\noindent
\begin{align}
    \bigg\lvert\frac{\delta x_n}{\delta x_1}\bigg\rvert = \exp{\left(\sum_{j=1}^{n-1} \lambda_{c_j}(x_j)\right)}.
    \label{eq:specChain}
\end{align}

\noindent For the expected behavior of stochastic systems where $m$ solution branches are chosen from with any weighted probabilities ($p_i, i\in[m], \sum p_i=1$), one can calculate the Lyapnov exponent by generalizing Eqs.~(\ref{eq:localLyap},\ref{eq:meanLyap},\ref{eq:theoryPDF}),

\noindent
\begin{gather}
    \lambda_i(x) = \log\left(|f_i'(x)|\right) \\
    \bar{\lambda} = \int_x \left(\sum_{i=1}^m p_i \lambda_i(x)\right) \rho_{\text{s.s.}}(x) dx \\
    \rho_{n+1}(x_{n+1}) = \sum_{i=1}^m p_i \frac{\rho_n(f^{-1}_i(x_{n+1}))}{|f_i'(f^{-1}_i(x_{n+1}))|}.\label{eq:recurGen}
\end{gather}

It is important to note that the expected mode localization of disordered Markov systems, while appearing qualitatively similar to the topologically protected edge modes of ordered systems, is \emph{not} topological in origin.
Expected ZM localization in disordered Markov systems comes from configurations with differing localizations being exponentially \emph{improbable}, not fundamentally \emph{impossible} (Fig.~\ref{fig:fig2}C).
For this reason, we introduce the term ``stochastic protection'' to refer to robust properties of disordered materials that are not necessarily topologically guaranteed, but are specifically exponentially unlikely to not occur.

\section*{Kink-antikink pairs and  design principles}
This stochastic protection of the exponential localization of the ZM to the generating end of the chain allows us to define the aforementioned concept of ``disorder-induced polarization.'' 
Furthermore, although the chain at large scales consistently exhibit a clear polarization towards the generating end (left), at smaller scales, we find isolated short sections of the chain where the ZM locally increases towards the right, displaying the opposite polarization (Fig.~\ref{fig:fig1}E). 
These sections give rise to kink-antikink pairs (following the notion of Ref.~\cite{zhou2017kink} for ordered chains), with $[-]$ and $[+]$ being localized quasi-topological charges.  

This mapping from small sections of the 1D chain exhibiting opposite polarization to kink-antikink pairs is supported by the low frequency spectrum of the system. 
As shown in Fig.~\ref{fig:fig3}A/B, a few modes exist between the ZM and the ``continuum'' (i.e., the counterpart of the bulk band in the ordered chain). 
The spatial profile of these ``soft modes'' reveal the nature of the kink-antikink pairs.  

This can be seen by computing the singular values and left and right singular space of the compatibility matrix $C$ and equilibrium matrix $Q=C^T$, define the linear mapping from rotor angles ($\theta$) to spring extensions ($e$), and the linear mapping from spring tensions ($t$) to rotor torques ($f$),

\noindent
\begin{equation}
    e_m = C_{mn}\cdot \theta_n  , \quad
    f_m = Q_{mn}\cdot t_m.
\end{equation}

\noindent
Singular value decomposition of $C$, a $(N-1)\times N$-dim matrix, reveals not only the ZM in its right null space, but also $N-1$ nonzero singular values $\omega_\alpha$, which are dimensionless frequencies that relate its left and right singular space

\noindent
\begin{align}
    C \theta_{\alpha} = \omega_\alpha e_\alpha . \nonumber\\
    Q e_\alpha = \omega_\alpha \theta_\alpha.
\end{align}

\noindent
Physically, this means that rotor rotation following the profile $\theta_{\alpha}$ causes extension profile described by $e_\alpha$ and the amplitude is proportional to $\omega_{\alpha}$. 
The identification of $\omega_{\alpha}$ as frequencies comes from the eigenvalues of $M^{-1}Q\cdot K \cdot C$ being true squared frequencies (where $M,K$ are the rotational inertia and spring constant matrices). 
Taking $M=K=I$ makes $\omega_{\alpha}$ these frequencies. 

For each low frequency singular value, we have a $C$ matrix right singular vector that is a ``quasi ZM,'' and a $C$ matrix left singular vector that is a ``quasi SSS.'' 
As shown in Fig.~\ref{fig:fig3}C, they are \emph{spatially apart}, with the quasi ZM coinciding with the max of the ZM and the quasi SSS coinciding with the min of the ZM. 
\emph{These quasi ZMs and SSSs sharply define the kink-antikink pairs ($[+]$ and $[-]$ charges) we mentioned above}. 
This is because kink-antikink pairs in 1D ordered Maxwell chains are defined as ZM and SSS interfaces~\cite{zhou2017kink}. 
Here we extend their meaning to include low frequency quasi ZMs and SSSs. 
The reason for why these quasi ZMs and SSSs coincide with the local maxima and minima of the topological ZM will be explained in the next section. 

The fact that configurations that defy the stochastically protected ZM localization of disordered rotor chains are not technically impossible, only improbable to realize by random chance, alludes to the possibility of a clever design principle that allows for the creation of designed systems with unique mechanical properties.
For specific configurations, to control ZM growth/decay for the next specific degree of freedom, one can always choose from the available local Lyapunov exponents (Eq.~(\ref{eq:specChain}), choosing a $\lambda_{i \in[m]}(x_n)$ for each $n$). 
However, this does not necessarily guarantee the availability of desired local Lyapunov exponents at each step needed to create desired steady-state behavior.
Here, we will demonstrate a simple design process to create different possible steady-state localizations, which can be combined to create any possible sequence of ZM growth/decay in a mechanical Markov system.

From our two solution functions $f_a, f_b$, we can construct two piecewise solution functions $f_\alpha, f_\beta$ with local Lyapunov exponents $\lambda_\alpha, \lambda_\beta$ respectively.
We want to create these piecewise functions such that successive choices from $f_\alpha$ ($f_\beta$) deterministically result in a negative (positive) Lyapunov exponent $\bar{\lambda}_\alpha$ ($\bar{\lambda}_\beta$) and therefore ZM decay (growth).
Finding closed-form solutions for the invariant probability density of a given iterating function is generally difficult \cite{lasota2013chaos}.
Therefore, knowing where to splice  $f_\alpha, f_\beta$ to necessarily and sufficiently create negative (positive) $\bar{\lambda}_\alpha ( \bar{\lambda}_\beta)$ is also generally difficult.
However, a merely \emph{sufficient} condition for $\bar{\lambda}_\alpha<0, \bar{\lambda}_\beta > 0$, is if there are absorbing subsets (subsets that will ``absorb'', as in contain indefinitely, all trajectories originating from any point in the domain) \cite{chueshov2002introduction} $u, v$ of the domain for $f_\alpha, f_\beta$ respectively, in which $|f'_\alpha(u)|<1, |f'_\beta(v)|>1$. 

In the case of the rotor chain system, this can be achieved by splicing $f_\alpha, f_\beta$ together at $\theta = \pi/2, 3\pi/2$ (see Fig.~\ref{fig:fig2}D), resulting in $u = 1.12\cup2.01$ (two stable fixed points), $v= [3.39, 6.03]$, $\bar{\lambda}_\alpha \simeq-0.570$ (the slopes at the two fixed points in $u$ are equal via symmetry, see SI Sec.~\ref{sym}), and $\bar{\lambda}_\beta \simeq +0.489$. 
By alternating between choosing solutions for subsequent rotor angles from $f_\alpha$ and $f_\beta$, we can construct a chain that alternates between ZM decay and growth in any desired sequential pattern (see Fig.~\ref{fig:fig2}E/F).
Such a process is possible for any 1D mechanical Markov system with at least two solution branches with appropriate absorbing subsets with all positive/negative slope. 

As this framework describes the behavior of systems that can be constructed via a Markov process, it applies naturally to self-assembled or synthesized materials where repeated additions of identical individual units need only be compatible with unit they are connecting to.
As these components can be added without a specific top-down design plan, it is common for them to realize structurally disordered or complex systems \cite{roach2022controlling}, making their ZM localization heretofore highly challenging to study.
For potentially controlling ZM localization in self-assembled systems, the simplicity in control over ZM growth/decay, as demonstrated in the rotor chain with a simple boolean ($f_\alpha$ or $f_\beta$), gives promising potential to simple and reliable customization in self-assembled systems.
In real self-assembled systems, hypothetical examples of possible methods to control this boolean include the presence/absence of a catalyst or magnetic field, or by having two different sets of components that attach to only one solution or the other.
The actual implementations of these control schemes are beyond the scope of the present work.

\section*{Perturbation Theory for Low-Frequency Displacement Modes}

\begin{figure*}[t!]
    \centering
    \includegraphics[width=\linewidth]{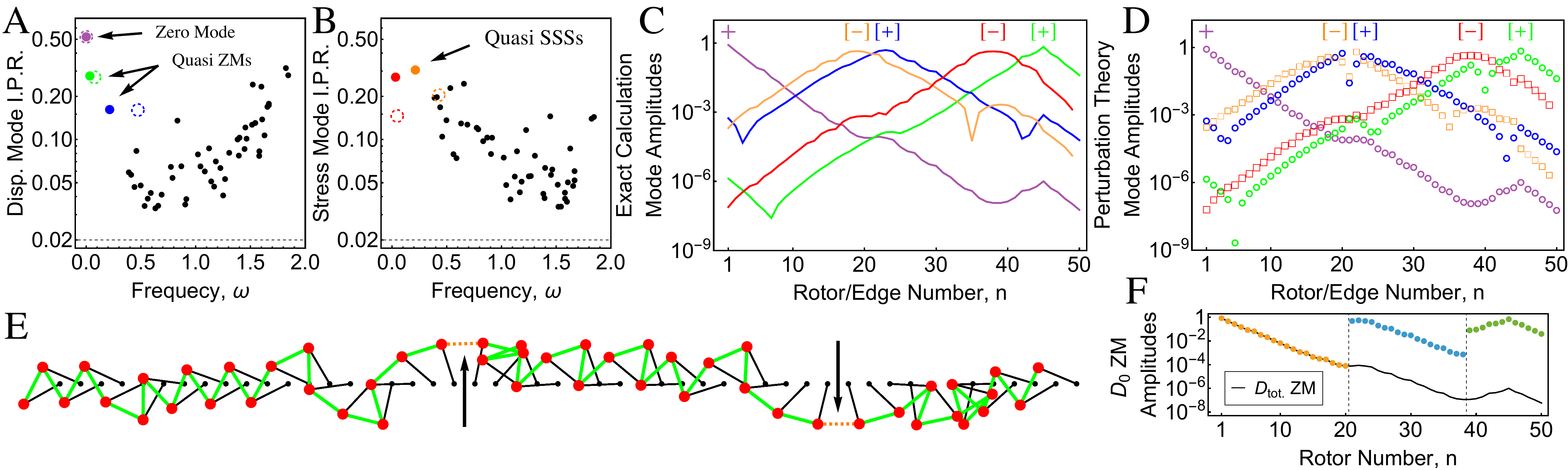}
    \caption{
    Perturbation theory explains these existence of qualitatively low-frequency modes and their localizations.
    (A) The displacement modes (solid points, right singular vectors of $C$, the compatibility matrix) of the system in Fig.~\ref{fig:fig1}E include the highly localized (high inverse participation ratio, I.P.R., $\sum_i x_i^4/\sum_i x_i^2$) topological zero-frequency mode (purple), as well as two relatively localized, qualitatively low-frequency modes (green, blue).
    First order perturbation theory results (dashed open circles) reasonably approximate these low-frequency modes.
    All modes are more localized than the usual I.P.R. of $~1/N$ of plane waves (horizontal dashed line).
    (B) The stress modes (solid dots, left singular vectors of $C$) include no topological states of self stress (no $\omega=0$, as expected by Maxwell counting), but do include two localized, low-frequency modes (orange, red).
    These frequencies are the same as low-frequency deformation modes, as the frequency is their shared singular value of $C$.
    (C) The right (left) singular vectors of $C$ that represent low-frequency displacement (stress) modes exponentially localize to local maxima (minima) of the topological ZM, all with roughly equal decay rates.
    (D) First order perturbation theory reproduces these localizations for both low-frequency displacement (circle markers) and stress (square markers) modes.
    (E) For displacement modes, perturbation theory is applied by removing edges 20 and 38, which are at ZM minima (orange, dashed) to obtain a zeroth order system.
    These removed edges are reintroduced as a perturbation.
    (F) The zeroth order system has three degenerate, non-overlapping ZMs ($\ket{\phi_j^{}}$, eigenstates of zeroth order Hamiltonian $D_0$), dashed lines indicate where edges were removed.
    }
    \label{fig:fig3}
\end{figure*}

As mentioned above, we find that positive (negative) quasi topological charges, defined at the local maxima and minima of the ZM, correspond exactly to qualitatively low-but-not-zero frequency displacement (stress) modes in the right (left) singular space of $C$ (Fig,~\ref{fig:fig3}A/B).
The modes exponentially localize to local maxima (minima) in the topological ZM, shown in Fig,~\ref{fig:fig3}C (note here that these springs have a finite stiffness $k=1$ while rotors are perfectly stiff).
These quasi ZMs and quasi SSSs, unique to disordered systems, are of great interest due to their close relation with soliton dynamics, as pairs of low-frequency modes appear at the middle and far-ends of defect regions. 
Here, we will explain the peak localization, decay rate, and frequency of displacement modes using perturbation theory (Fig.~\ref{fig:fig3}D).
Similar perturbation theory for stress modes can be found in the SI Sec.~\ref{perForSSS}.

To the system in Fig.~\ref{fig:fig1}E, from each rotor with a locally minimal ZM amplitude, we will remove an edge (such as edges 20 and 38, shown in Fig.~\ref{fig:fig3}E).
This will be our $0$th order system.
It consists of three completely independent shorter rotor chains, each with a ZM of frequency exactly 0 ($\ket{\phi_{j}^{}}$, $j \in [3]$), shown in Fig.~\ref{fig:fig3}F.
Since we removed edges adjacent to minimal ZM amplitudes, these new sub-chains have no quasi states of self-stress and each only a single ZM.
Our $0$th order Hamiltonian will then be (using our original compatibility matrix $C$)

\noindent
\begin{align}
    D_0 = C^T \cdot K_{0} \cdot C,
\end{align}

\noindent with a spring constant ($k=1$) matrix $K_0$ of

\noindent
\begin{align}
    (K_0)_{i,j} \,= \begin{cases}
        1,&   \text{if }\, i=j \text{ and } i \notin \text{removed edges} \\
        0,& \text{otherwise}. \\
    \end{cases} 
    \label{eq:kMat}
\end{align}

We now apply degenerate perturbation theory (to be justified later), where the previously removed springs will be added back in as a perturbation Hamiltonian matrix $D_1$,

\noindent
\begin{align}
    D_1 &= C^T \cdot K_{1} \cdot C, \\
    (K_1)_{i,j} &\,= \begin{cases}
         1,&   \text{if }\, i=j \text{ and } i \in \text{removed edges} \\
        0,& \text{otherwise}, \\
    \end{cases} 
\end{align}

\noindent
(where we have gone ahead and taken $\lambda$, the expansion parameter, to 1).
The resulting eigenvalues (eigenvectors) of the three by three matrix,

\noindent
\begin{align}
    \bra{\phi_\mu^{(0)}} D_1 \ket{\phi_\nu^{(0)}},
\end{align}

\noindent
are the first-order energy shifts $E_i^{1}$ to the ``good'' basis states $\ket{\phi_i^{(0)}}$ (coefficients of linear combination of $\ket{\phi_j^{}}$ that gives $\ket{\phi_i^{(0)}}$).
For our example system of three sub-chains, this procedure splits our original three degenerate ZMs into two low-frequency modes and one exactly-zero mode, with frequencies similar to those of the exact solutions from before, as shown in Fig.~\ref{fig:fig3}A.

We also calculate the first-order corrections to these states by using the unperturbed basis vectors $\ket{\psi_k^{(0)}}$ outside of our degenerate subspace ($S$),

\noindent
\begin{align}
    \ket{\phi_i^{(1)}} = \sum_{k \notin S} \frac{-1}{E_k} \bra{\psi_k^{(0)}}D_1\ket{\phi_i^{(0)}} \ket{\psi_k^{(0)}}, \label{eq:pertstate}
\end{align}

\noindent
to which we add to our original ``good'' basis states $\ket{\phi_i^{(0)}}$ and renormalize.
These resulting first order states are shown in Fig.~\ref{fig:fig3}C/D to be in good qualitative agreement with the exact results.

To justify the use of perturbation theory here, we must find that the expansion coefficients from Eq.~(\ref{eq:pertstate}) are small compared to the unperturbed non-zero energies:

\noindent
\begin{align}
    \big|\bra{\psi_k^{(0)}}D_1\ket{\phi_i^{(0)}}\big| \ll \big|E_k^{(0)}\big|. \label{eq:pertcon}
\end{align}

\noindent
$(D_{1})_{n,m}$ is only non-zero when rotor $n$ and $m$ are both incident to the same ``perturbation'' spring. 
Each $\ket{\psi_k^{(0)}}$ is not necessarily expected to have low amplitude at sites $n, m$, so the condition in Eq.~(\ref{eq:pertcon}) is entirely dependent on the degenerate ZM $\ket{\phi_i^{(0)}}$ (which are each a linear combination of the non-overlapping $\ket{\phi_j}$) having very low amplitude at the site of the perturbation spring.
This is precisely why we removed springs at ZM minima, creating $\ket{\phi_j}$ with small amplitudes at cites $n$ and $m$.

As this condition for perturbation theory here is increasingly met ($\ket{\phi_j}$ has smaller amplitude at boundaries), the energy corrections become vanishingly small, explaining why the green mode (which strongly meets this condition) from Fig.~\ref{fig:fig3}A has a smaller low-frequency than the blue mode (which moderately meets this condition).
Similarly, as this condition increasingly holds, perturbation corrections to the degenerate states (Fig.~\ref{fig:fig3}F, which necessarily have the same decay rates as the ZM) become increasingly small, explaining why the decay rates of of these low-frequency modes are similar to the Lyapunov-exponent-derived decay rates of topological ZMs.
It is significant that these low-frequency displacement modes robustly localize to ZM amplitude maxima with similar decay rates to the ZM, as driving these systems with experimentally realistic low-but-not-exactly-zero frequency will not result in drastically different behavior.

\section*{Lyapunov Exponents for Explaining Soliton Propagation}

As discussed in relation to Fig.~\ref{fig:fig1}D/E, we find that anomalous soliton dynamics occurs in ``defect'' regions around quasi topological charges defined in the previous section.
This includes either slow soliton propagation where each intermediate rotor completes nearly a full $2\pi$ rotation, or instant soliton teleportation where each intermediate rotor hardly rotates.
We will now establish why these defects appear and why they are fundamentally related to quasi topological charges.

Consider a rotor chain in the linkage-limit (for each rotor/spring, $k\rightarrow \infty$). In this case, the angle of each rotor can only change by activating the single ZM: $|\dot{\theta}_n/\dot{\theta}_m| = |(\delta\theta_n/\delta t)/(\delta\theta_m/\delta t)|$.
We can use Eq.~(\ref{eq:specChain}) for a given set of solution choices ($c_j = a \text{ or } b; j \in [n-1]$) to show that rotor angular velocities also follow a Lyapunov exponent relationship,

\noindent
\begin{align}
    \bigg\lvert\frac{\dot{\theta}_n}{\dot{\theta}_m}\bigg\rvert &= e^{\lambda_{n,m}(\theta_m)}= \exp{\left(\sum_{j=m}^{n-1} \lambda_{c_j}(\theta_j)\right)},
    \label{eq:timeDep}
\end{align}

\noindent
Here, $\lambda_{n,m}(\theta_m)$ is the ``relative Lyapunov exponent'' of $\theta_n$ to $\theta_m$, equivalent to $\log|d\theta_n/d\theta_m|$.
From Eq.~(\ref{eq:timeDep}), we can simulate the dynamics of rotors chains by directly updating a ``driving'' rotor over one time step, then compatibly updating the angles of all subsequent rotors with appropriate scale factors such that kinetic energy is conserved. 
Further details regarding simulation methodology can be found in SI Sec.~\ref{dynamics}.

\begin{figure*}[t!]
    \centering
    \includegraphics[width=\linewidth]{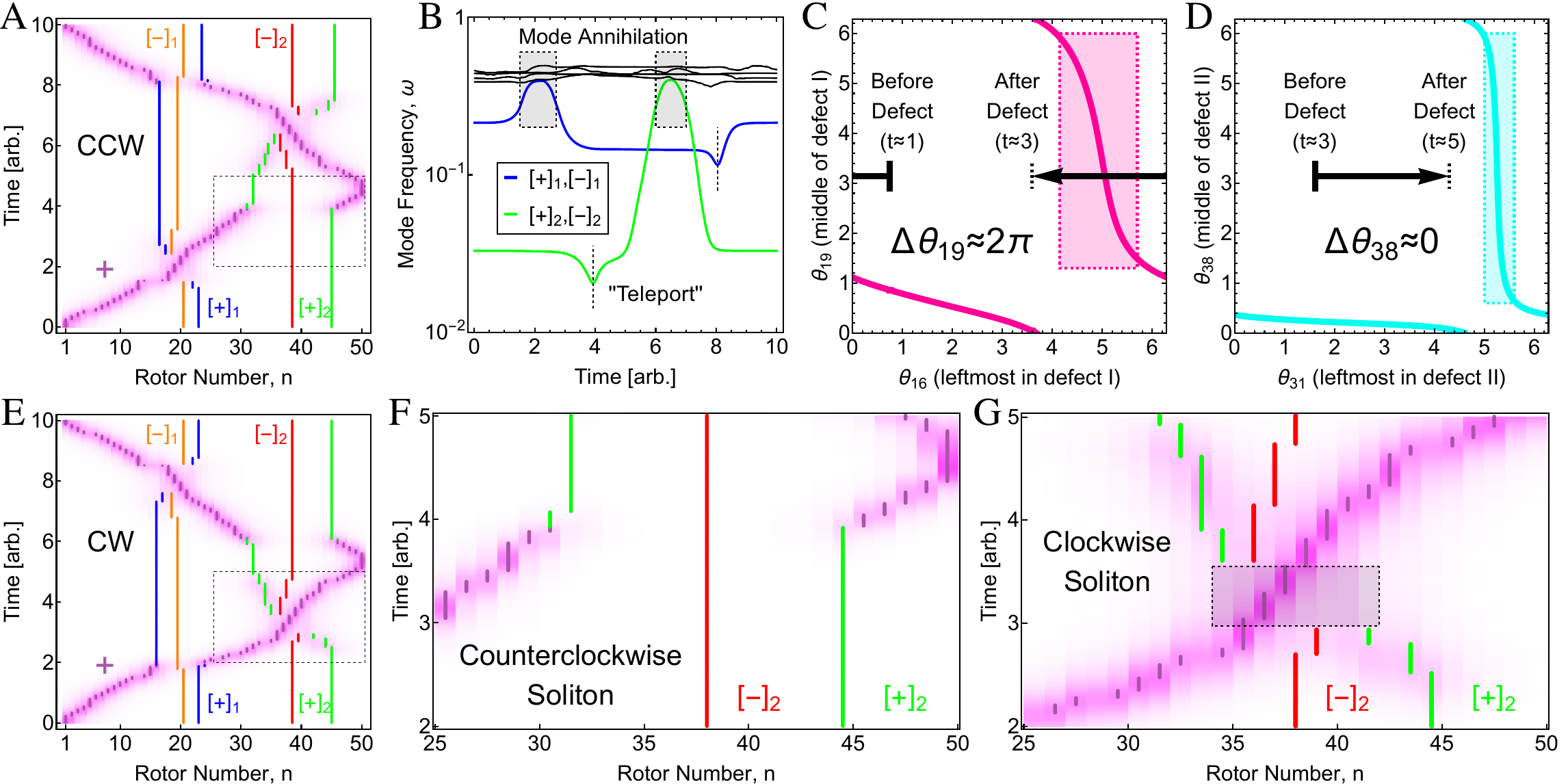}
    \caption{
    Regular soliton transport in disordered rotor chains is interrupted at ``defects,'' where soliton transport is either slow or an instant teleport. 
    (A) Transport of a counterclockwise soliton (pink hue $\propto |\dot{\theta}|$, same as Fig.~\ref{fig:fig1}A) with and overlay of locations of the quasi topological charges (rotor with highest amplitude of corresponding singular vector).
    Each ``defect'' region is comprised of a positive ([$+$]) and negative ([$-$]) quasi topological charge, which here are listed with the subscript of the defect they correspond to.
    (B) The frequencies of quasi ZMs and quasi SSSs from (A), note that coupled charges of the same defect have the exact same frequency due to sharing the same singular value of $C$.
    During ``slow transport'' (shaded boxes), quasi topological charges annihilate, raising the frequency of these usually ``low'' frequency modes to be on par with other ``high'' frequency modes (next four lowest shown in black).
    (C/D) The behavior of a defect (slow or instant soliton transport) is determined by the domain of angles swept out by the leftmost angle ($\theta_{16}$ for I and $\theta_{31}$ for II) during transport.
    If the leftmost angle traverses through a rapid cycle (narrow shaded region of angles in the leftmost rotor in which a middle rotor completes nearly a complete rotation), the defect will have slow transport (C).
    If the leftmost angle does not traverse through a rapid cycle, the defect will have instant transport (D). 
    (E) Transport of a \emph{clockwise} soliton on the exact same system.
    Here, the behavior of defects is opposite from the counterclockwise case in panel A (slow $\leftrightarrow$ instant) because the compliment set of angles is swept out by $\theta_{16}, \theta_{38}$, negating the inclusion of a rapid cycle.
    (F/G) Zoomed in panels of the dashed boxes in A/E with shaded box to show mode annihilation. 
    }
    \label{fig:fig4}
\end{figure*}

Assuming that there is no defect at the beginning of a rotor chain, as rotor 1 initially rotates, $\lambda_{2,1}$ increases. 
$\lambda_{2,1}$ will necessarily eventually become positive at a later time, as both $f_a, f_b$ have regions of their domain with a slope of magnitude greater than 1.
This results in rotor 2 having a higher angular velocity than rotor 1, becoming the location of the soliton as the rotor with the highest angular velocity.
Rotor 2 will now rotate until $\lambda_{3,2}$ necessarily becomes positive, and so on.
This pattern will continue through any region of the rotor chain with uniform effective polarization.
However, this pattern gets interrupted when the soliton is at a rotor $m$, and a distant rotor $n$ has a greater relative Lyapunov exponent $\lambda_{n,m}$ than any rotor between $n$ and $m$.
In this case, rotor $n$ will initially rotate more than $m+1$, explaining why defects occur between distant rotors with a relative Lyapunov exponent ($\lambda_{n,m}$) of approximately zero (equivalent to the condition $\delta\theta_n \approx \delta\theta_m$ introduced in Fig.~\ref{fig:fig1}).

For a distant rotor $n$ to have a higher Lyapunov exponent relative to a given rotor $m$ than rotor $m+1$ has relative to $m$, there must necessarily be a region of increasing $\lambda_{i,m}$ with respect to $i$, for some range of $i \in [m+1,n]$.
This is exactly a region of opposite disorder-induced polarization.
Assuming that this region of opposite polarization does not extend to the end of the chain, there must be at least one local minima and one local maxima in the ZM amplitude between rotors $m$ and $n$.
These extrema are exactly quasi topological charges (also low frequency stress and displacement modes) from the previous section, explaining why we always find quasi topological charges in regions of defective soliton transport.

Having now established why Lyapunov exponents explain the locations of defects (rotors between $n,m$ when $\lambda_{n,m} \approx0$), and how defects are intrinsically linked to quasi topological charges, we now explain why these defect regions produce the two distinct types of modified transport and how these quasi topological charges interact with the transporting soliton.
There are two types of modified transport, one where soliton velocity is decreased and each intermediate rotor completes nearly a full rotation (``slow transport''), and one with nearly instantaneous soliton transport and each intermediate rotor hardly rotates (``instant transport'').
Interestingly, during slow transport, the positive and negative quasi charge associated with the defect in question ``annihilate'' with each other, while during instant transport, the positive quasi charge teleports in the opposite direction of the true topological charge (Fig.~\ref{fig:fig4}A/B).
The former fact should not be totally surprising, as each rotor in the bulk of a defect having very high angular velocity during slow transport corresponds to large relative Lyapunov exponent of these rotors with respect to rotor $m$, eliminating the local minimum/maximum pair in this region. 

Returning to the fact that defects are the result of a distant rotor moving more than a local one ($\lambda_{n,m} \approx0$), one may wonder why all modified soliton transport is not an instant transport, as in all cases of modified transport the soliton \emph{does}, at least briefly, instantly send some energy from rotor $m$ to $n$.
However, the range of values that $\theta_m$ must sweep out in order to complete this soliton transport to rotor $n$ does not necessarily leave all intermediate rotors $k \in (m,n)$ unchanged.
The amount that any intermediate rotor $k$ rotates by can be expressed as

\noindent
\begin{align}
    \Delta\theta_k = \int_{\theta_m(t_i)}^{\theta_m(t_{\text{f}})} \bigg| \frac{d\theta_k}{d\theta_m}\bigg| d\theta_m = \int_{\theta_m(t_i)}^{\theta_m(t_{\text{f}})} e^{\lambda_{k,m}(\theta_m)} d\theta_m,
\end{align}

\noindent
where the function $\big|\frac{d\theta_k}{d\theta_m}\big|$ has a slope with a magnitude of approximately zero over most of it's domain with a small region of very large magnitude slope where $\theta_k$ completes nearly a full rotation, as demonstrated in Fig.~\ref{fig:fig4}C/D.
Therefore, if transporting a soliton from rotor $m$ to $n$ requires $\theta_m$ to pass through one of these ``narrow rapid cycles'' in the function $\theta_k(\theta_m)$, the soliton will have to wait for the each intermediate rotor to complete nearly a full cycle before traveling to rotor $n+1$.
Each of these sequential full cycles requires the conserved total kinetic energy to be mostly concentrated into the single rapidly rotating rotor, making this process take a long time, hence the transport being ``slow.''
The existence of these rapid cycles is related to the fact that $\lambda_{k,m} - \lambda_{n,m} < 0$, where the further negative this quantity is (ie: larger defect), the more narrow the rapid cycle (in $\theta_m$ space).
For further discussion related to the relation between $\lambda_{k,m}$ and the existence of rapid cycles, see SI Sec.~\ref{rapidcycles}.

We can now see why returning solitons necessarily have oppositely modified transport. 
As $\theta_{n+1}(\theta_n)$ is monotonic, any angles swept out by an incoming soliton (moving right to left) are mutually exclusive with the angles by the outgoing (moving left to right) soliton.
Therefore, if a narrow rapid cycle is traversed by the outgoing soliton, it cannot be traversed by in the incoming soliton.
Similarly, solitons of opposite chirality ($\theta_1$ rotating clockwise or counter-clockwise) must experience opposite modified transport, as changing the sign of $\dot{\theta}_1$ necessarily changes the sign of all $\dot{\theta}$ from the monotonicity of $\theta_{n+1}(\theta_n)$.
If an initially clockwise soliton would have caused $\theta_m$ to traverse from $\theta_{m,t_i}$ to $\theta_{m,t_{\text{f}}}$ along a path with a narrow rapid cycle, an initially counter-clockwise soliton would then have to traverse between those angles in the other direction in $\mod{2\pi}$ space, avoiding the narrow rapid cycle.
The differences between outgoing/incoming and clockwise/counter-clockwise solitons can be seen in Fig.~\ref{fig:fig4}A/E and their magnified views in Fig.~\ref{fig:fig4}F/G.
For discussion relating soliton chirality through the lens of spin-orbit coupling, see SI Sec.~\ref{soCouple}.

Because low frequency modes have significantly increased frequency during slow transport (quasi topological charges get annihilated) and decreased frequency during instant transport (as shown in Fig.~\ref{fig:fig4}B), we show how one can predict the type of modified soliton transport analytically without simulating any dynamics by using the Hellmann-Feynman theorem (SI Sec.~\ref{hft}).
In combination with using relative Lyapunov exponents to identify the \emph{locations} of defects, one can completely predict the time-dependent behavior of a disordered rotor chain using only parameters known to its static configuration.
This one-way chirality-dependent transport is highly unusual for in mechanical systems in general, and is even unusual in rotor chains as it appearance is dependent on intrinsic disorder.
Here, we have made major progress towards analytically understanding the complex dynamics of these disordered systems.

\section*{Conclusion}
We report a new class of disorder-induced wave phenomena in 1D using a simple mechanical Markov chain model. 
By identifying the Lyapunov exponent via a space-to-time mapping as the key quantity governing localization, we demonstrate that disorder---rather than destroying coherent behavior---induces a form of stochastic polarization: ZMs are exponentially unlikely to delocalize away from the one particular end of the chain depending on the order of construction.
This provides a powerful predictive tool for both disordered and designed systems, and allows us to utilize the concept of Lyapunov exponents to engineer arbitrary localization patterns, a functionality important for many engineering problems~\cite{liu2000locally,tsakmakidis2007trapped,zhang2018fracturing,xiu2022topological}. 

Beyond the linear regime, we show that disorder gives rise to rich nonlinear dynamics that has not been observed before, including defect-controlled soliton transport and teleportation phenomena that are sensitive to chirality. 
These behaviors highlight how disorder can serve as a design principle, generating responses unattainable in ordered systems while remaining analytically tractable.

More broadly, the framework we establish here suggests new pathways for harnessing disorder for novel wave phenomena. 
The Markov-process perspective naturally applies to self-assembled or bottom-up synthesized systems, where a large class of systems follow particle-by-particle attachment type of growth, resembling the Markov construction~\cite{de2017holistic,jun2022classical}. 
In these systems, stochasticity is unavoidable yet may be exploited for robust function. 
Future work may extend these ideas to higher dimensions, to active or non-Hermitian systems~\cite{ding2022non} where Lyapunov spectra carry additional richness, and to experimental platforms where disorder can be precisely engineered~\cite{mao2024complexity}. 
Taken together, our results point to a new paradigm in which Lyapunov exponents provide a systematic language to design, predict, and control the emergent properties of disordered  systems.

{\it Acknowledgments.}---The authors thank Daniel Duffy and Nicholas Boechler for helpful discussions. The project was supported in part by the Office of Naval Research (MURI N00014-20-1-2479) and the National Science Foundation Center for Complex Particle Systems~(Award \#2243104). 

\clearpage

\section*{Supplementary Information}

\subsection{Fixed point slopes from symmetries} \label{sym}

Multiple times in the main text we mention that the slopes of $f_a$, $f_b$ at the fixed points of those function are equal.
To be more specific, there are four fixed points in total, two of them have a slope with magnitude less than 1 (attractive), two of them have slopes with magnitude greater than 1 (repulsive).
The two attractive fixed points have equal slopes and the two repulsive fixed points have equal slopes, and furthermore, these two different slopes are reciprocal of each other.
We will now explain why these must be true via symmetries of these functions.

Because our coordinate labeling is two periodic, but our physical system is not, our map of $\theta_n$ to $\theta_{n+1}$ must remain unchanged if we slide the system one rotor to the right or left.
This real-space translation would cause our coordinate labeling to change $\theta_n$ to $\pi - \theta_n$.
In state-space, this is rotation of $\pi$ radians around the point $(\pi/2,\pi/2)$ or $(3\pi/2,3\pi/2)$, consequently these maps must remain unchanged under these rotation around these points.
Rotating the stable fixed points around $(\pi/2,\pi/2)$ and the unstable fixed points around $(3\pi/2,3\pi/2)$ returns the two stable and unstable fixed points respectively without changing the slope.
Therefore, the two stable fixed points and the two unstable fixed points have the same slopes.

Similarly, if we were to instead measure each odd-index rotor \emph{clockwise} from the positive vertical and each even-index \emph{counter-clockwise} from the negative vertical, we would have changed nothing about the system, but each coordinate would be changed from $\theta_n$ to $2\pi - \theta_n$.
This transformation corresponds to a mirror reflection across the line $\theta_{n+1}=2\pi-\theta_n$.
Under this reflection, the two stable fixed points get mapped to the unstable fixed points and visa-versa.
This map inverts the slopes, making the unstable and stable fixed points have slopes that are reciprocal of each other.
This fact should be totally unsurprising, as the ZM decay rate of a Kane Lubensky chain and the ZM growth rate of that chain's left-to-right mirror image should be opposite.

\subsection{Analytic forms of $f_\alpha$, $f_\beta$}\label{analytic}

In the main text, having not yet introduced $f_\alpha, f_\beta$ by the time we introduce $f_a, f_b$, we mention for simplicity that the analytic forms of the latter are included in the Supplementary Information.
However, the default analytic solutions from our computer algebra solver were $f_\alpha$ and $f_\beta$, so those are the forms presented here.
Of course, in the reverse way that we argued on could construct $f_\alpha, f_\beta$ in the main text, one could use these forms to construct $f_a, f_b$.

\begin{align}
    f_{\alpha(\beta)} = \arctan\left(\frac{y_{\alpha (\beta)}}{x_{\alpha (\beta)}}\right)\mod 2\pi,
\end{align}

\noindent where shared terms in $y, x$ do not necessarily cancel, as the quadrant of the point $(x,y)$ is taken into account.

\begin{align}
    y_{\alpha} = \Bigg(r^2 \left(3 a^2-l^2\right) \sin (\theta_n)+\sqrt{z} \nonumber\\
    +a r (a-l) (a+l)-a r^3 (\cos (2 \theta_n)-3) \nonumber\\+2 r^4 \sin (\theta_n)\Bigg)/\Big(2 r^2 \left(a^2+2 a r \sin (\theta_n)+r^2\right)\Big)
\end{align}

\begin{align}
    x_{\alpha} = \Bigg(\sec (\theta_n) (a+r \sin (\theta_n)) \sqrt{z} \nonumber\\
    -r^3 \cos (\theta_n) (a^2+2 a r \sin (\theta_n)-l^2 \nonumber\\
    +2 r^2)\Bigg)/\Big(2 r^3 \left(a^2+2 a r \sin (\theta_n)+r^2\right)\Big)
\end{align}

\begin{align}
    y_{\beta}=\Bigg(r^2 \left(3 a^2-l^2\right) \sin (\theta_n) -\sqrt{z} \nonumber\\
    +a r (a-l) (a+l)-a r^3 (\cos (2 \theta_n)-3)+ \nonumber\\
    2 r^4 \sin (\theta_n)\Bigg)/\Big(r^2 \left(a^2+2 a r \sin (\theta_n)+r^2\right)\Big)
\end{align}

\begin{align}
    x_{\beta} = -\Bigg(\sec (\theta_n) (a+r \sin (\theta_n)) \sqrt{z} \nonumber\\
    +r^3 \cos (\theta_n) \left(a^2+2 a r \sin (\theta_n)-l^2+2 r^2\right)\Bigg)\nonumber\\
    /\Big(r^3 \left(a^2+2 a r \sin (\theta_n)+r^2\right)\Big)
\end{align}

\begin{align}
    z =-r^4 \cos ^2(\theta_n) \Big(2 r^2 \left(a^2-2 l^2\right)+\left(a^2-l^2\right)^2 \nonumber\\
    +2 a r (2 (a-l) (a+l) \sin (\theta_n)-a r \cos (2 \theta_n))\Big)
\end{align}

As a reminder, these solutions come from solving:

\begin{align}
    a^2 + 2r(r+r\cos{(\theta_{n}+\theta_{n+1})} + \\ \nonumber a\sin{(\theta_n)} - a \sin{(\theta_{n+1})} ) = l^2,
\end{align}

\noindent which is likely significantly easier to copy into a computer algebra solver of the reader's choice than it is to manually copy the solutions above.

\subsection{Derivation of recursion relation for probability densities}\label{rhos}

Here, we will derive Eqs.~(\ref{eq:theoryPDF}, \ref{eq:recurGen}), which are essentially applications of Ruelle–Perron–Frobenius operators \cite{viana2016foundations} for probabilistic maps.
We will start from probability change of variables for a single function $x_{n+1} = f(x_n)$ between scalars $x_n,x_{n+1}$:

\begin{align}
    \big|\rho_{n}(x_{n}) dx_{n}\big| &= \big|\rho_{n+1}(x_{n+1}) dx_{n+1}\big| \\
    \rho_{n+1}(x_{n+1}) &= \rho_{n}(x_{n}) \bigg|\frac{dx_{n}}{dx_{n+1}}\bigg| \\
    \rho_{n+1}(x_{n+1}) &= \frac{\rho_{n}(x_{n})}{\big|f'(x_{n})\big|}\\
    \rho_{n+1}(x_{n+1}) &= \frac{\rho_{n}(f^{-1}(x_{n+1}))}{\big|f'(f^{-1}(x_{n+1}))\big|}.
\end{align}

Of course, the step between line two and three will need to be changed to a weighted sum if there are multiple functions that relate a differential $dx_{n+1}$ and $dx_n$:

\begin{align}
    \rho_{n+1}(x_{n+1}) &= \sum_{i=1}^mp_i\frac{\rho_{n}(x_{n})}{\big|f'_i(x_{n})\big|} \\
    \rho_{n+1}(x_{n+1}) &= \sum_{i=1}^m p_i \frac{\rho_n(f^{-1}_i(x_{n+1}))}{|f_i'(f^{-1}_i(x_{n+1}))|},
\end{align}

which is the form shown in Eq.~(\ref{eq:recurGen}).

\subsection{Discrete Markov transition matrix approximation for $\rho_\text{s.s.}$}\label{disMarkov}

As mentioned in the main text, one can approximate the steady-state probability density by constructing a discrete Markov transition matrix $\mathcal{M}$ \cite{lasota2013chaosMARKOVMATRIX}, which discretizes the state space of rotor angles into $\eta$ ``bins'', and maps the probability that an angle at site $n$ in bin $\zeta \in [\eta]$ (where we are using the notation $[\eta]$ to denote the set $\{1,2,3,\dots,\eta\}$) will result in an the angle at site $n+1$ being in bin $\chi \in [\eta]$.
The matrix $\mathcal{M}$ will then have $\eta$ rows and columns, with elements defined as follows,

\noindent
\begin{align}
    \mathcal{M}_{\zeta,\chi} \,= \begin{cases}
        p_a,&   \text{if}\, f_a(2\pi\zeta) \in [2\pi (\chi-1)/\eta,2\pi \chi/\eta] \\
        p_b,& \text{if}\, f_b(2\pi\zeta) \in [2\pi (\chi-1)/\eta,2\pi \chi/\eta]\\
        0,&   \text{otherwise.}\\
    \end{cases} 
    \label{eq:markovMat}
\end{align}

\noindent
As our rotor chain system is disordered and permits solutions $f_a, f_b$ that are bijective, our ``dynamical system'' is trivially ergodic, and therefore has only one stationary (or steady-state) probability density \cite{viana2016foundations}.
This unique stationary density will be an eigenmode of $\xi$ of $\mathcal{M}$ that has an eigenvalue equal to $1$ (all others $<1$).
The elements of $\xi$ then represent the steady-state probability distribution,

\noindent
\begin{align}
    \xi_\chi = p(\theta_{\text{s.s.}} \in [2\pi (\chi-1)/\eta,2\pi \chi/\eta]).
    \label{eq:markovEig}
\end{align}

\subsection{Dynamic simulation methodology}\label{dynamics}

To simulate the dynamics of rotor chain systems, we begin with a completely static rotor chain configuration, shown in Fig.~\ref{fig:fig4}A.
For each rotor, we store which solution (as a function of the previous rotor) it was chosen to be (ie: $f_a, f_b$).

At this initial time, we find the relative Lyapunov exponent of each rotor with respect to the ``driving'' rotor, which here, was chosen to be rotor 1, $\lambda_{n,1}$.
From these relative lyapunov exponents, we can find the velocities of all rotors with respect the velocity of the driving rotor (here, rotor 1):

\begin{align}
    \dot{\theta}_n = \dot{\theta}_1 e^{\lambda_{n,1}}.
\end{align}

\noindent The total kinetic energy of the chain is then obviously $K \propto \dot{\theta}_n \cdot \dot{\theta}_n = \dot{\theta}_1^2 \sum(e^{\lambda_{n,1}})^2$.

We can now update our driving rotor by some small quantity $\delta\theta_{1,t=1}$, such that $\theta_{1,t=2} = \theta_{1,t=1} + \delta\theta_{1,t=1}$.
We calculate this small change in angle such that total kinetic energy is conserved by:

\begin{align}
    \delta\theta_1 \propto \sqrt{\frac{\dot{\theta}_1^2}{K}} = \frac{1}{\sqrt{\sum(e^{\lambda_{n,1}})^2}} = \frac{1}{\text{Norm}(e^{\lambda_{n,1}})}.
\end{align}

\noindent The proportionality constant for the proportional relationship above determines the speed scale of the soliton transport.

Importantly, we do not update the remaining rotors in a similar way.
If we did update the other rotors through calculating some $\delta\theta_n$ from Lyapunov exponents, each spring would be slightly extended or compressed after each time step.
As the system is in the linkage-limit, we know that no springs can be stretched, but that also means that all $\theta_n$ remain solutions of the functions that they were chosen to be solutions of initially.
At each time step, we simply reconstruct the chain, using the stored sequence of solution choices ($c_n$):

\begin{align}
    \theta_{n+1,t} = f_{c_n}(\theta_{n,t}).
\end{align}

\noindent From these new rotor angle solutions, we can calculate the new relative Lyapunov exponents, and repeat the process indefinitely.

\subsection{Lyapunov exponents for tensions}

\begin{figure}[ht]
    \centering
    \includegraphics[width=\linewidth]{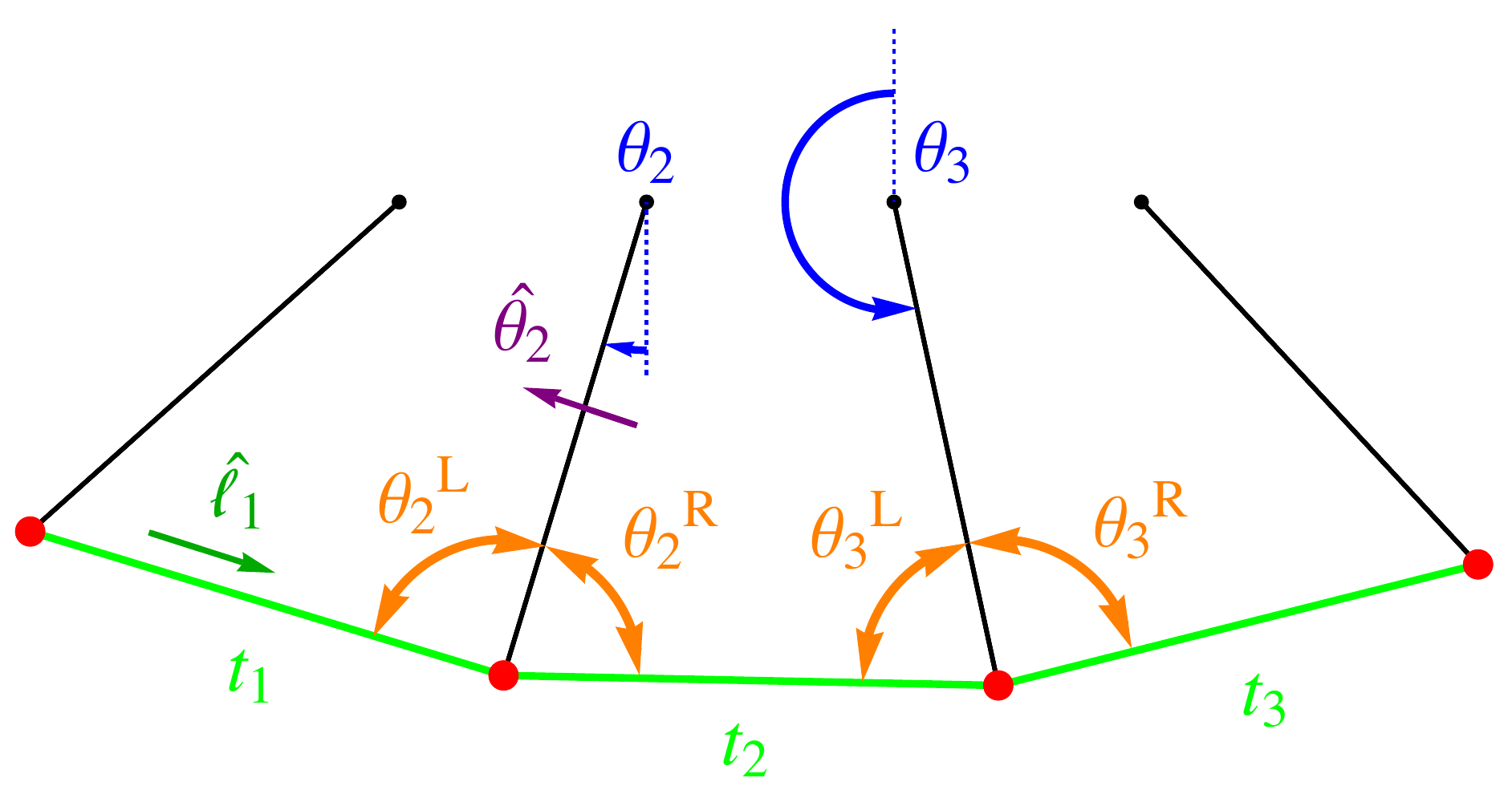}
    \caption{
    Diagram of angles and vectors used in the derivation of SSS Lyapunov exponents.
    }
    \label{fig:figSI}
\end{figure}

Consider a variation of the rotor chain system, where the rotors on the two spatial ends of the chain are pinned in place. 
By adding these two constraints, this ``pinned'' chain now has no ZM and a now a SSS.
By allowing rotor arms to support any force, we can find a recursion relationship for nearest-neighbor spring tensions throughout the pinned chain via torque balance around the rotor between the two springs:

\noindent
\begin{align}
    t_n \hat{l}_n \cdot \hat{\theta}_{n+1} &= t_{n+1} \hat{l}_{n+1} \cdot \hat{\theta}_{n+1} \nonumber \\
    \frac{t_{n+1}}{t_n} &= \frac{\hat{l}_n \cdot \hat{\theta}_{n+1}}{\hat{l}_{n+1} \cdot \hat{\theta}_{n+1}}.\label{eq:tensionRecur}
\end{align}

\noindent
$\hat{\theta}_{n+1}$ is a trivial function of $\theta_{n+1}$, and $\hat{l}_i, \hat{l}_{n+1}$ can also be found as functions of only $\theta_n, \theta_{n+1}, \theta_{n+2}$.
Of course, with defined solution choices ($a, b$), both $\theta_{n+1}, \theta_{n+2}$ are simply functions of $\theta_n$, allowing us to write the right-hand side of Eq.~(\ref{eq:tensionRecur}) as only a function of $\theta_n$:

\noindent
\begin{align}
    \frac{t_{n+1}}{t_n} &= g_{ij}(\theta_n), i \text{ and } j = a \text{ or }b.
\end{align}

While this recursion relationship is not explicitly a recursion relationship between infinitesimal changes in a trajectory (like $\delta\theta$), we can still define a Lyapunov exponent here to measure the growth/decay of tensions:

\noindent
\begin{align}
    \frac{t_{n+1}}{t_n} &= e^{\lambda_{ij}(\theta_n)}, i \text{ and } j = a \text{ or }b.
\end{align}

\noindent
Similar to how we found the expected growth/decay of ZM amplitude in the main text, here we can find expected SSS growth/decay by averaging over four possible solution pairs:

\noindent
\begin{align}
    \EX{\frac{t_{n}}{t_1}} &= e^{\bar{\lambda}(n-1)} \\
    \bar{\lambda} = \int_0^{2\pi}&\frac{1}{4}\left(\lambda_{aa}(\theta)+\lambda_{ab}(\theta)+\lambda_{ba}(\theta)+\lambda_{bb}(\theta)\right) \rho_{\text{s.s.}}(\theta)d\theta.
\end{align}

\begin{figure}[ht]
    \centering
    \includegraphics[width=\linewidth]{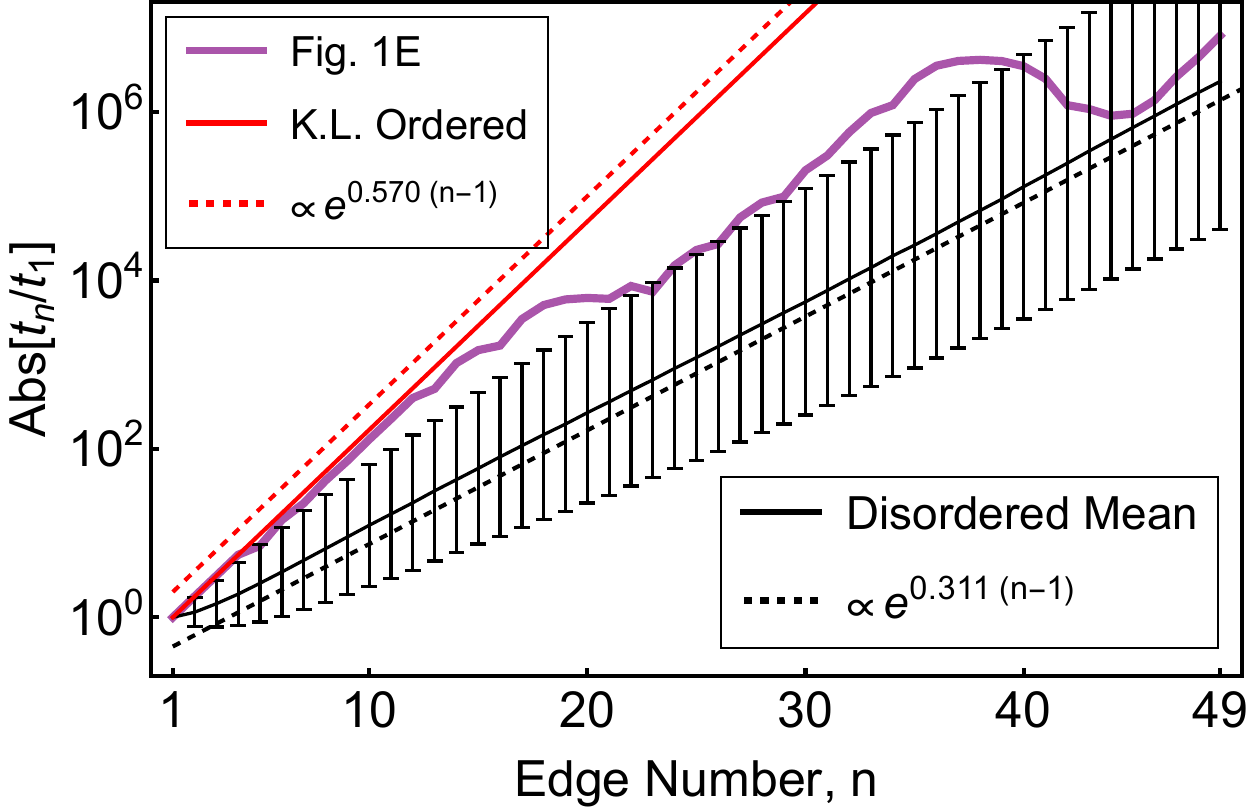}
    \caption{
    Lyapunov exponents for SSS in pinned rotor chains.
    }
    \label{fig:figSI2}
\end{figure}

\noindent
For our usual geometry ($l=1, r=1, a=0.5$), this calculation yields $\bar{\lambda} \simeq +0.311$, which is exactly opposite of the ZM Lyapunov exponent, as far as our discrete $\rho_{\text{s.s.}}$ can tell.
Similarly, for a ``pinned'' KL ordered chain, we find a SSS with $\bar{\lambda} \simeq +0.570$, which is exactly opposite of the ordered ZM Lyapunov exponent (see Fig.~\ref{fig:figSI2}).

The steady-state Lyapunov exponents for SSS being the opposite of Lyapunov exponents for ZM is not a coincidence.
Using angles defined in Fig.~\ref{fig:figSI}, we can find two related recursion relations for tension and displacements:

\noindent
\begin{align}
    t_{n+1} \sin(\theta_i^R) &= t_n \sin(\theta_i^L) \\
    \delta\theta_{n+1} \sin(\theta_{n+1}^L) &= \delta\theta_n \sin(\theta_n^R),
\end{align}

\noindent
which give:

\noindent
\begin{align}
    \frac{t_{n+m}}{t_n} &= \frac{\prod_{k=n}^{n+m-1}\sin(\theta_k^L)}{\prod_{k=n}^{n+m-1}\sin(\theta_k^R)} \\
    \frac{\delta\theta_{n+m}}{\delta\theta_n} &= \frac{\prod_{k=n}^{n+m-1} \sin(\theta_k^R)}{\prod_{k=n+1}^{n+m}\sin(\theta_k^L)} \\
    \implies \frac{t_{n+m}}{t_n}\frac{\delta\theta_{n+m}}{\delta\theta_n} &= \frac{\sin(\theta_n^L)}{\sin(\theta_{n+m}^L)}, \label{eq:tensfmrel}
\end{align}

\noindent
showing that these two ratios are nearly reciprocal with some finite offset, which is independent of length.
In the large-system-size and steady-state limits, this finite offset becomes increasing negligible, giving nearly exactly opposite Lyapunov exponents.

\subsection{Perturbation theory for low-frequncy stress modes} \label{perForSSS}

In the main text, we show perturbation theory results for stress modes that approximate the low-frequency stress modes of the \emph{unpinned} (regular) rotor chain.
Here, having now shown the relationship between edge tensions and rotor ZM amplitude (Eq.~(\ref{eq:tensfmrel})), we can explain and justify our procedure for these perturbation theory results.

Similar to the unpinned rotor chain's topological ZM, the pinned rotor chain's topological SSS has two pairs of local minima and maxima (obvious pair in defect II, subtle pair in defect I).
If we wanted to approximate the \emph{pinned} system's low-frequency stress modes, we could follow a completely analogous procedure used to approximate the unpinned system's low-frequency displacement modes.
That is, exclude some component from $H_0$ that creates degenerate non-overlapping exactly zero-frequency modes (here, that would constitute pinning rotors, before it constituted cutting edges).
Then, reinstating that component in $H_1$.
For this specific pinned system, that would require pinning rotors 22 and 45 (in addition to the already pinned rotors 45 and 50), and allowing 22 and 45 to rotate again as a ``perturbation''.
This would yield two low-frequency stress modes (like those we want to explain in the unpinned system), but also a topological SSS localized to edge 49, which is not present in the unpinned system.

To use perturbation theory to explain the low-frequency stress of the unpinned system, all rotors we ``pin'' in $H_0$ must all be allowed to rotate as a perturbation, to prevent a resulting topological SSS.
Because relative Lyapunov exponents are associative, we know that the degenerate SSS of any $H_0$ for the unpinned chain will have the same slopes as the single topological SSS of the pinned chain (which is shown in Fig.~\ref{fig:figSI2}).
We will want to ``join'' these degenerate modes where there amplitude is low (to meet the analogous perturbation theory criteria established previously for displacement modes), so we should ``split'' these modes (pin rotors) at all local minima of the pinned chain's topological SSS.
For our example system, that constitutes pinning rotors 1, 22, and 45, but \emph{not} 50. 
We then allow all of 1, 22, 45 to rotate again as a perturbation.
Perturbation theory results from this process are shown in Fig.\ref{fig:fig3}.

\subsection{Existence of rapid cycles in defects}\label{rapidcycles}

\begin{figure}[ht]
    \centering
    \includegraphics[width=\linewidth]{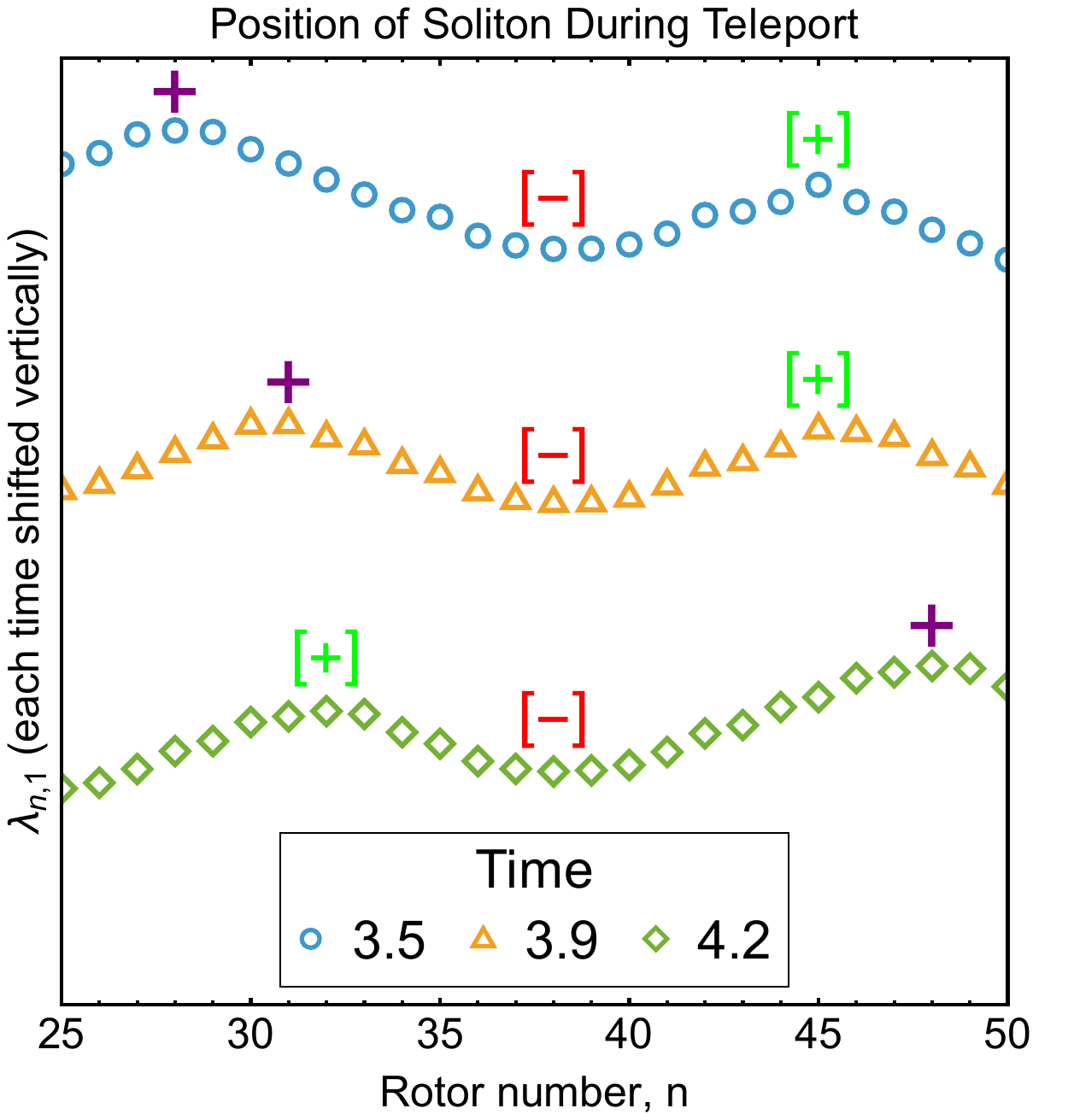}
    \caption{
    Locations of zero-frequency soliton (purple +) and low frequency displacement/stress modes ($[+],[-]$) during soliton teleportation. Soliton teleportation corresponds to \emph{not} passing through a rapid cycle.
    }
    \label{fig:figSI3}
\end{figure}

\begin{figure}[ht]
    \centering
    \includegraphics[width=\linewidth]{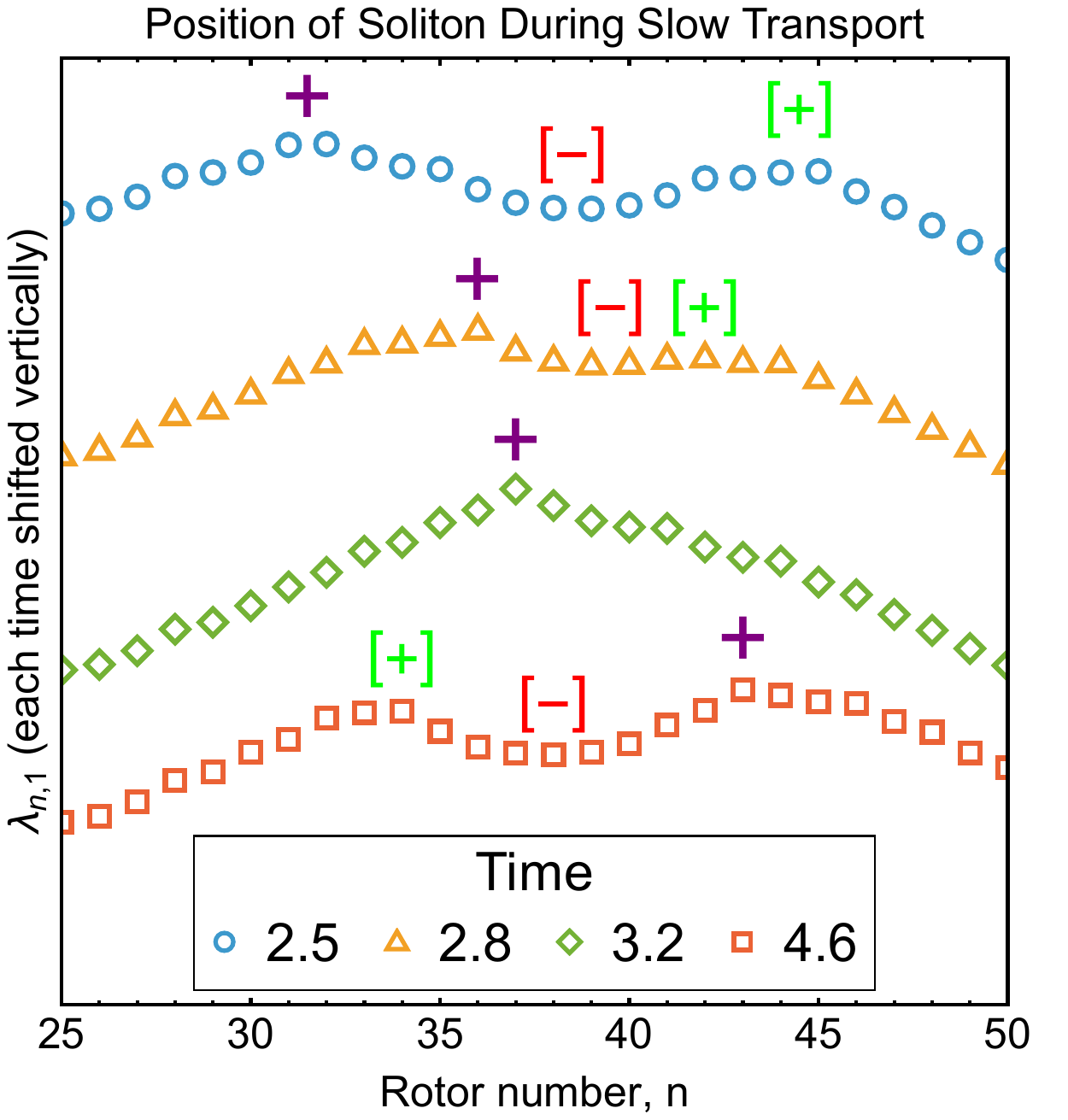}
    \caption{
    Locations of zero-frequency soliton (purple +) and low frequency displacement/stress modes ($[+],[-]$) during slow soliton transport. Slow soliton transport corresponds to passing through a rapid cycle, causing each rotor in the bulk of the defect to complete nearly a full rotation.
    }
    \label{fig:figSI4}
\end{figure}

It is worth discussing the conditions that lead to the existence of narrow rapid cycles in $\theta_k(\theta_m)$, as they are key to modified soliton transport.
Consider only defects that, based on their solutions choices, are expected to have exponentially decaying ZM amplitude followed by exponentially growing ZM amplitude.
This condition is clearly met for defects I and II, as shown by the decay rates in Fig.~\ref{fig:fig1}C, however, highly unusual defects with multiple minima/maxima pairs would fail this condition.
For some rotor $k$ in a defect between rotors $m,n$, we will then have,

\noindent
\begin{align}
    g_k(\theta_m) &\equiv\bigg|\frac{d\theta_k}{d\theta_m}\bigg|\\ 
     \implies\EX_{\theta_m}\log\left(g_k(\theta_m) \right)&= \EX_{\theta_m}\lambda_{k,m}(\theta_m) \label{eq:logg}\\ 
     &\sim \begin{cases}
        \bar{\lambda}(k-m),&   \text{if } k \leq (m+n)/2\\
        \bar{\lambda}(n-k),&  \text{if } k > (m+n)/2,\\
    \end{cases} 
\end{align}

\noindent
where we are using the subscript $\theta_m$ to denote an expectation value over a uniform distribution of $\theta_m$ (as opposed to a uniform distribution over some other space).
Here, we are taking an expectation value over $\theta_m$ because we are interested in a property of the function $\theta_k(\theta_m)$.
$\EX_{\theta_m}\lambda_{k,m}(\theta_m)$ is then increasingly negative the closer $k$ is to the middle of the defect.
By expanding $\log\left(g_k(\theta_m) \right)$ around $g_k(\theta_m) = 1$, we get the following series (where we apply linearity of expectation),

\begin{align}
    \EX_{\theta_m}\log\left(g_k(\theta_m) \right)&=\EX_{\theta_m}(g_k-1) -\frac{1}{2}\EX_{\theta_m}(g_k-1)^2 \nonumber\\
   & + \frac{1}{3} \EX_{\theta_m}(g_k-1)^3 - \frac{1}{4}\EX_{\theta_m}(g_k-1)^4 + ... \label{eq:expansion}
\end{align}

Because the function $\theta_{n+1}(\theta_n)$ is a smooth bijective function between $\theta_n$, $\theta_{n+1}$ for all $n$, completing a cycle in $\theta_n$ (wrapping around [0,2$\pi$] space back to a starting angle) necessitates completing exactly one cycle of $\theta_{n+1}$.
By induction, completing a cycle of any rotor necessitates completing a cycle of \emph{all} rotors, giving,

\noindent
\begin{align}
    \frac{1}{2\pi}\int_0^{2\pi}\bigg|\frac{d\theta_k}{d\theta_m}\bigg|d\theta_m &= 1 \\
    \EX_{\theta_m}\left(\bigg|\frac{d\theta_k}{d\theta_m}\bigg|\right) &=1,
    \label{eq:condition}
\end{align}

\noindent
Therefore, the first term of Eq.~(\ref{eq:expansion}) is zero.

The following terms of Eq.~(\ref{eq:expansion}), each being proportional to $(g_k-1)^i$, are proportional to the higher order \emph{moments} of the distribution of $g_k(\theta_m) = |d\theta_k/d\theta_m|$,

\begin{align}
    \EX_{\theta_m}\log\left(g_k(\theta_m) \right)&=0 -a \cdot \text{Variance}(g_k) \nonumber\\
    &+ b \cdot\text{Skewness}(g_k) - c \cdot \text{Kurtosis}(g_k) + ... 
\end{align}

\noindent
We can now see that the value of $\EX_{\theta_m}\log\left(g_k(\theta_m) \right)$, which was just the expected relative Lyapunov exponent of intermediate rotors (Eq.~(\ref{eq:logg})), is a measure of how \emph{nonuniform} the distribution of slopes of $\theta_k(\theta_m)$ is.
A rapid cycle in $\theta_k(\theta_m)$ is precisely a manifestation of this high non-uniformity.
Admittedly, a highly nonuniform distribution of slopes does not preclude the possibility of a defect having two rapid half-cycles, or three rapid third-cycles, and so on, but the authors have never observed this phenomena in defects with single minima/maxima pairs.

For the effects of these rapid cycles/teleports on the relative Lyapunov exponents of each rotor in a defect, see Figs.~\ref{fig:figSI3},\ref{fig:figSI4}.

\subsection{Soliton chirality and spin-orbit couples}\label{soCouple}

It may be instructive to understand the chirality and soliton velocity under the lens of spin-orbital coupling.    
We define a total rotor spin angular momentum, $\vec{s}=\hat{z}\sum_{i} L_i$, (where, by coordinate convention, $L_i = \dot{\theta} _i (-1)^i$).
The \emph{direction} (sign) of $\vec{s}$ is necessarily conserved by the monotonicity of $\theta_{n+1}(\theta_n)$, however the value of this quantity is not necessarily conserved due to each $L_i$ being defined relative to a different reference point (each rotor pivot point).
The  momentum of the soliton $\vec{p}=\dot{\vec{X}}$ (which we call ``orbital momentum'' following the literature of spin-orbital coupling of waves) where $X$ is the soliton center, is also not conserved, as pinning points of the rotors exert forces.

In an ordered KL chain, spin and orbital momentum are decoupled: kicking the rotors CW or CCW results in the same soliton velocity.  In the mechanical Markov chain we study here, disorder induces a spin-orbital coupling.  Remarkably, this spin-orbital coupling is concentrated at the  defects (kink-antikink pairs).  In particular, each defect $a$ carries a dipole coupling coefficient $\Delta_a \hat{y}$, determining the coupling strength $\Delta_a \hat{y} \cdot (\vec{s} \times \vec{p})$. This captures the main feature in Fig.~\ref{fig:fig4}, where region I shows slow transport when $\vec{y}\cdot(\vec{s} \times \vec{p}) >0$ and teleport when $\vec{y}\cdot(\vec{s} \times \vec{p}) <0$, while region II is opposite. 

\subsection{Hellmann-Feynman theorem for defect type identification} \label{hft}

As stated in the main text, low frequency modes have significantly increased frequency during slow transport (quasi topological charges get annihilated) and decreased frequency during instant transport (as shown in Fig.~\ref{fig:fig4}B).
Smooth continuous changes in rotor angles require that the frequency of low-frequency modes can only change smoothly and continuously.
Therefore, by finding the sign of the time derivative of that frequency leading up to a defect can reveal what the transport the system will undergo within the defect (positive $\rightarrow$ slow, negative $\rightarrow$ instant teleport).
In fact, we empirically observe that the sign of this derivative is positive/negative if the \emph{next} defect will have slow/instant transport regardless of how ``close'' the soliton is to that defect (Fig.~\ref{fig:fig4}B).
Of course, the sign of time derivative of the frequency (square root of the energy $E_j$) of some low-frequency mode $\ket{\psi_j}$ can be found numerically by simply taking a single time-step forward, but it can also be found analytically with only the initial configuration of the system through the application of the Hellmann-Feynman theorem,

\noindent
\begin{align}
    \frac{dE_j}{dt} = \sum_{i=1}^{N} \frac{d\theta_i}{dt} \bra{\psi_j}\frac{\partial D}{\partial \theta_i}\ket{\psi_j},
\end{align}

\noindent which similarly predicts that opposite chirality solitons will experience opposite transport as taking $\frac{d\theta_i}{dt} \rightarrow -\frac{d\theta_i}{dt} \implies \frac{dE_j}{dt} \rightarrow -\frac{dE_j}{dt}$.

\bibliography{apssamp}

\end{document}